\documentclass[11pt, a4paper]{article}

\usepackage{times}
\normalfont
\usepackage[T1]{fontenc}
\usepackage{amsmath,amstext,amsgen,amsbsy,amsopn,amsfonts,amssymb}
\usepackage{graphics}
\usepackage{graphicx}
\usepackage{epsfig}
\usepackage{wrapfig, lipsum, booktabs}
\usepackage{bm}
\usepackage{color}
\usepackage{algorithmic}
\usepackage{algorithm}
\usepackage{hyperref}
\usepackage{esvect}
\usepackage{multirow}
\usepackage{array}

\usepackage{caption}
\usepackage{sidecap}

\usepackage{titlesec}
\usepackage{lipsum}% just to generate text for the example
\titlespacing*{\section}
{1pt}{3.0ex plus 1ex minus .2ex}{1.1ex plus .2ex}
\titlespacing*{\subsection}
{1pt}{2.5ex plus 1ex minus .2ex}{1.0ex plus .2ex}

\begin{document}
%%\begin{frontmatter}                           % The preamble begins here.

%\pretitle{Pretitle}
\title{Clustering of Transcriptomic Data for the Identification of Cancer Subtypes\footnote{The final publication is available at IOS Press through DOI \url{10.3233/978-1-61499-927-0-387})}}
%%\runningtitle{IOS Press Style Sample}
%\subtitle{Subtitle}

\author{
Xiaochun Chen$^{\dag}$, Honggang Wang$^{\dag}$, Donghui Yan$^{\ddag}$
\vspace{0.1in}\\
$^\dag$Department of Electrical and Computer Engineering\\ University of Massachusetts Dartmouth, MA\vspace{0.05in}\\
$^\ddag$Department of Mathematics and Program in Data Science\\ University of Massachusetts Dartmouth, MA\\[0.05in]
%%\\
}

%%\date{Aug 18, 2018}
\date{}
\maketitle

\begin{abstract}
\noindent
Cancer is a number of related yet highly heterogeneous diseases. Correct identification of cancer subtypes is critical for clinical decisions. 
The advance in sequencing technologies has made it possible to study cancer based on abundant genomics and transcriptomic (-omics) 
data. Such a data-driven approach is expected to address limitations and issues with traditional methods in identifying cancer subtypes. 
We evaluate the suitability of clustering--a data mining tool to study heterogenous data when there is a lack of sufficient understanding of 
the subject matters--in the identification of cancer subtypes. A number of popular clustering algorithms and their consensus are 
explored, and we find cancer subtypes identified by consensus clustering agree well with clinical studies. 
\end{abstract}

\section*{Introduction}
Cancer is a number of related diseases involving unusual cell growth with the potential to invade or spread to other 
parts of the body \cite{Bertino2002}. Every year there are about 9 million of deaths, or about 1/6 of all deaths, due to 
cancer, according to the World Health Organization (WHO) \cite{who}. %%\url{http://www.who.int/news-room/fact-sheets/detail/cancer}). 
Among challenges in cancer treatment are the heterogeneity and the evolving nature of cancer. According to the US 
National Cancer Institute (NCI) \cite{nciCancer}, %%\url{https://www.cancer.gov/about-cancer/understanding/what-is-cancer}), 
there are over 100 cancer subtypes. Genetic mutations, cancer microenvironmental, immune, and therapeutic selection pressures 
all dynamically contribute to tumor heterogeneity. Heterogeneity may lead to cells with differential molecular signature 
within a single tumor tissue or varying levels of sensitivity in treatment. In addition, cancer heterogeneity also contributes 
to therapy resistance. Therefore, deciphering cancer heterogeneity will have major impact to cancer treatment, the 
development of effective therapy \cite{DagogoShaw2018}, and personalized medicine strategy \cite{RosenthalMcGranahan2017}.
\\
\\
The accurate identification of tumor subtypes can also facilitate clinical implementations \cite{BijlsmaSadanandam2017}. 
However, traditional methods are typically based on a single index and are often clinically unsatisfactory. For example, 
the WHO groups breast cancers into 17 subtypes based on histopathological staging, but such a classification has little 
impact on clinical decisions \cite{KourouExarchos2015}.
Most gene expression based approaches in gastro--intestinal tumors identify a mesenchymal subgroup with a poor 
prognosis. Other approaches include immune cell profiling, which is considered prognostic but not widely used due to the 
lack of consistency of genes \cite{Lyons2017ImmuneCP}. It is thus desirable to use additional clues or to combine multiple 
criteria in identifying cancer subtypes. 
\\
\\
The advent of affordable genomic and transcriptomic methods has made it possible to adopt a data-driven approach to 
identify cancer subtypes \cite{CamastraDTS2015}. Due to the lack of sufficient understanding of cancers, clustering 
appears to be an ideal tool for identifying cancer subtypes with biological and clinical significance. Clustering also makes 
it possible to combine multiple criteria whenever relevant data are available. The advance in high throughput sequencing 
technology has made it possible to use genetic information to assist the diagnosis of cancers. The Cancer Genome Atlas (TCGA) 
\cite{CGARN2014, TCGA2015} published by the US National Institutes of Health consists of comprehensive maps of key 
genomic and transcriptomic changes in major types of cancer. As cancers are caused by errors in DNA for cells to 
grow abnormally thus the abundant information in TCGA 
could be used to help find cancer subtypes. In this paper, we will assess the suitability of various clustering methods 
for such a task. This will provide a basis for further work in using TCGA or other relevant data in identifying cancer subtypes 
with a data-driven approach. As publications on TCGA mostly use non-consensus clustering algorithms such as K-means
clustering, hierarchical clustering and spectral clustering, we focus more on consensus clustering, and in particular a comparison 
of consensus and non-consensus algorithms. 
\\
\\
The remainder of this paper is structured as follows. In Section~\ref{section:methods}, we will briefly describe several popular 
clustering algorithms used in this study. Then we present our experiments and results in Section~\ref{section:exp}.
We conclude in Section~\ref{section:conclusion}.
\section{Methods}
\label{section:methods}
Clustering is an important method of data mining and scientific discovery. It aims to partition a set of data such that data points 
within the same cluster are ``similar" while points from different clusters are dissimilar. A number of different clustering methods 
have been used in the clustering of cancer -omics data. This includes K-means clustering \cite{ZhangLiu2013}, spectral clustering 
or its variants \cite{MaZhang2017, YangMichailidis2016}, similarity network fusion (SNF) \cite{SNF2014}, latent variable analysis 
\cite{MoWangSeshan2013} etc. However, very few studies use consensus clustering for -omics data. In this 
section, we briefly review several popular clustering algorithms which we will use for performance comparison on the clustering 
of cancer-omics data. This includes the classical K-means clustering and hierarchical clustering, the more recent spectral 
clustering, and consensus clustering (aka cluster ensemble).
\subsection{K-means clustering}
$K$-means clustering \cite{lloyd1982} is a simple clustering algorithm and remains one of the most popular 
clustering algorithms for over half a century. Starting with a set of randomly selected initial cluster centers, the algorithm 
alternates between two steps: assign all the points to its nearest cluster center (the center of 
mass of all points in the cluster), and recalculate the new cluster centers, and stops when there are no further changes 
on the cluster centers. More details can be found in \cite{lloyd1982, KaufmanRousseeuw1990}. 
\subsection{Hierarchical clustering}
Hierarchical clustering is a class of clustering algorithms that first organize the data in a tree-like hierarchy 
(i.e., {\it dendrogram}), and then form clusters by cutting the dendrogram at a certain height. This causes 
the tree to be split into several branches, and data in each branch form a cluster. The hierarchy can be formed 
bottom-up or top-down. For bottom-up, it starts by treating each data point as a singleton cluster (i.e., a cluster 
consisting of only one data point) and then keeps on merging the the ``nearest" clusters until all the data belong 
to a single final cluster. In top-down, initially all data belong 
to one cluster. Then, recursively, the cluster with largest diameter is divided until all clusters are singletons. 
Hierarchical clustering is attractive due to its tree-like representation of the data, which can be easily visualized. 
More details can be found in \cite{KaufmanRousseeuw1990, Ward1963}. 
\subsection{Spectral clustering and SNF}
Spectral clustering \cite{ShiMalik2000, NgJordan2002} is widely acknowledged as the method of choice for 
clustering, due to its superior empirical performance. It 
has been successfully applied in a wide range of applications such as computer vision, web search, social 
network mining, and market research etc. It works on an affinity graph over data points $X_1,...,X_N$, and 
seeks to find a ``minimal" graph cut. The affinity graph uses its edges to encode the similarity between pairs 
of points, $X_i$ and $X_j$; the edge weights, $a_{ij}$, form the affinity matrix $A = (a_{ij})_{i,j=1}^N$. Then 
it computes the Laplacian of $A$ defined as 
\begin{equation*}
\mathcal{L}(A)=D^{-1/2} (D-A)D^{-1/2},
\end{equation*}
where $D$ is a diagonal matrix with its i-th diagonal being the sum of similarity of point $X_i$ to all other points. 
The cluster membership is derived by a further K-means clustering of eigenvectors of $\mathcal{L}(A)$ \cite{NgJordan2002} 
or by a procedure that recursively looks at the the sign of the second last eigenvector of the Laplacian of the similarity 
matrix of data points in the sub-clusters \cite{ShiMalik2000}. SNF \cite{SNF2014} combines data, possibly under 
different metrics, into a single metric and then uses spectral clustering to group the data points. 
%%\\
%%\\
%%
%%
%%
%%\\
\subsection{Consensus clustering}
Issues with single clustering algorithms include the choice of number of clusters, and the lack of tools in assessing the 
confidence of clustering results. Consensus clustering \cite{StrehlGhosh2002} readily addresses these issues in a 3-step
procedure. First, it generates multiple instances of clustering with random restarts, resampling \cite{MontiTamayo2003}, 
randomized feature pursuit \cite{CF}, or deep representation and pseudo labelling \cite{LiuShaoFu2016,LiuShaoLiFu2017}. 
Then it combines these multiple clustering instances to form a consensus matrix. The $(i,j)$-element of the consensus 
matrix defines the similarity between data point $X_i$ and $X_j$. It is proportional to the frequency that these two points 
are in the same cluster in the ensemble. Typically, a clear block diagonal shape (sharp contrast between diagonal blocks 
and non-diagonal elements) of the consensus matrix indicates a `good' clustering, but it is preferable to use it together with 
other measures (as no single measure captures the full characteristics of clustering). The third step derives clustering membership 
from the consensus matrix, by any clustering method that accept inputs in the form of a similarity matrix such as spectral 
clustering. Consensus often generates a better clustering than non-consensus clustering, and is less sensitive to noise or 
outliers in the data. We will explore consensus for K-means, hierarchical clustering and SNF. 
It is possible to form a consensus of these three, which we leave to future work.

\section{Experiments}
\label{section:exp}
Our experiments use data from TCGA. TCGA consists of information that maps key genomic changes to a dozens of major 
cancer types. It hosts data collected from over 11,000 cancer patients with high throughput sequencing technologies. In this work, 
we analyze the RNA-seq and miRNA-seq datasets of the TCGA stomach adenocarcinoma (STAD) \cite{CGARN2014} for cancer 
subtype identification. The RNA-seq and miRNA-seq data consists of expression levels, in numerical scale, of over 20000 genes. 
We evaluate the performance of K-means clustering, hierarchical clustering, SNF, and their consensus. 

\subsection{Data collection and preprocessing}
The TCGA STAD datasets are downloaded with the R package TCGA2STAT \cite{WanAllenLiu2016}. We then merge these 
datasets and remove the missing values. To effectively reduce the data dimensionality, we choose the genes by their variance---those 
with larger variances are chosen. The idea is essentially that of principal component analysis \cite{Hotelling1933}, except that here the 
components are the expression level along different genes. We choose top 5000 genes from RNA-seq, and top 500 from miRNA-seq. 
Such a choice is supported by the fact that the chosen genes carry over 99\% of the variance in the data.

\subsection{Evaluation metric}
As there is no gold standard for measuring cancer subtypes, we use two metrics for the evaluation of clustering performance. 
The first is the average silhouette width (ASW) \cite{Rousseeuw1987}. The silhouette width of a data point is the difference 
between the average of its distance to the neighboring cluster and that to the same cluster, normalized by the max of the two. 
For a given clustering assignment, the ASW is calculated as follows. For a data point $X_i$, let $a(i)$ be the average distance 
of point $X_i$ to other points in the same cluster, and $b(i)$ be the average distance of $X_i$ to all points in the ``closest" cluster
that it does not belong to. Then the silhouette width of $X_i$ is calculated as 
\begin{equation*}
s(i) = \frac{b(i)-a(i)}{max(a(i),b(i))}.
\end{equation*}
The ASW is the average silhouette width of all data points. ASW takes value in the range $[-1,1]$, and the larger the better 
clustering. The second metric is the consensus matrix. These are among the popular metrics for cluster quality measure. 
Other measures such the clustering accuracy \cite{YanHuangJordan2009,XingNgJordanRussell2002} is not applicable here as 
the true labels are unknown.

\subsection{The number of clusters}
\begin{figure}[h]
\centering
\begin{center}
\hspace{-0.15in}
\includegraphics[scale=0.27,clip]{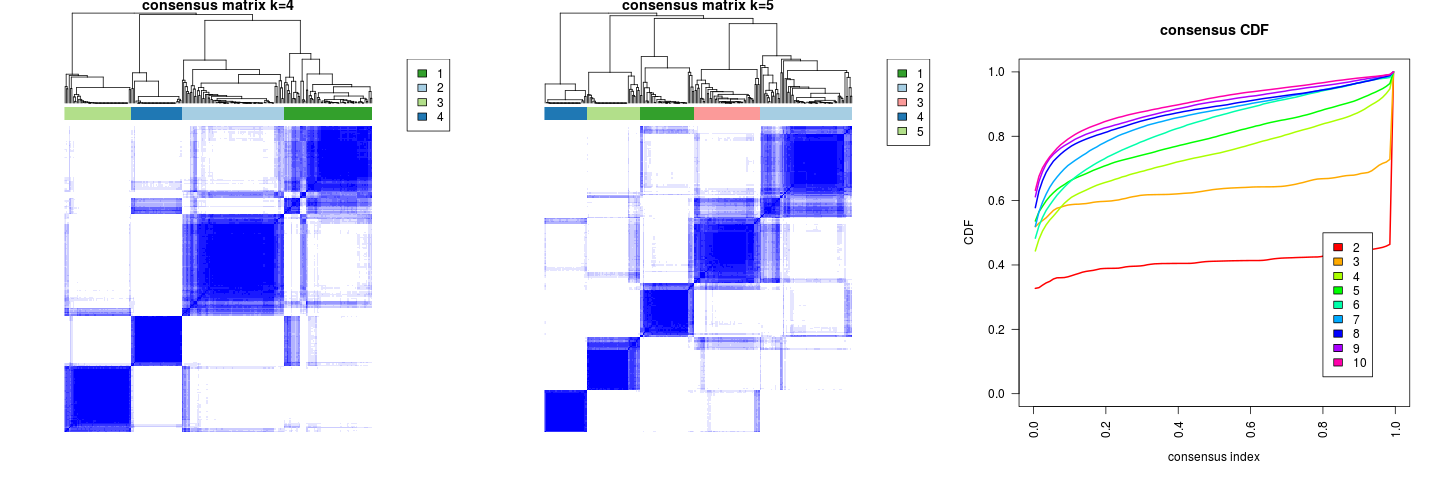}
\end{center}
\abovecaptionskip -5pt
\caption{\it Consensus matrix and consensus CDF by consensus clustering over SNF.} 
\label{figure:fig1}
\end{figure}

The number of clusters, K, is determined by consensus clustering over SNF. A range of different values from the set $\{2,3,4,5,6\}$ are 
searched, and Figure~\ref{figure:fig1} shows the consensus matrix under different K's (only $K=4,5$ are shown to save space). 
Based on a visual inspection of the consensus matrix, the number of clusters is set to be 4. The consensus CDF curve on the right 
further confirms this (the curve for $K=4$ is fairly flat for consensus indices in the range of $[0.1,0.9]$). This is also consistent with 
the literature \cite{BijlsmaSadanandam2017}. Here CDF is the percentage of pairs of data points that have a consensus index up 
to a given value, in analogy to the cumulative distribution function in probability; the consensus index of a pair of points is 
the percentage of times they are in the same cluster out of the number of times they are both included in a consensus instance. 
A flat CDF curve in the range of $[0.1,0.9]$ reflects the belief that, under a good consensus clustering algorithm, points in the same 
``natural" cluster would meet frequently among consensus clustering instances and rarely otherwise. 

\subsection{Clustering results}
\begin{figure}[h]
\centering
\begin{center}
\hspace{0cm}
\includegraphics[scale=0.68,clip]{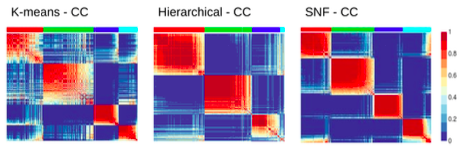}
\end{center}
%%\abovecaptionskip -32pt
\caption{\it Consensus matrix for consensus K-means clustering, hierarchical clustering and SNF.} 
\label{figure:fig2}
\end{figure}

\begin{figure}[htp]
\centering
\begin{center}
\hspace{0cm}
\includegraphics[scale=0.23,clip]{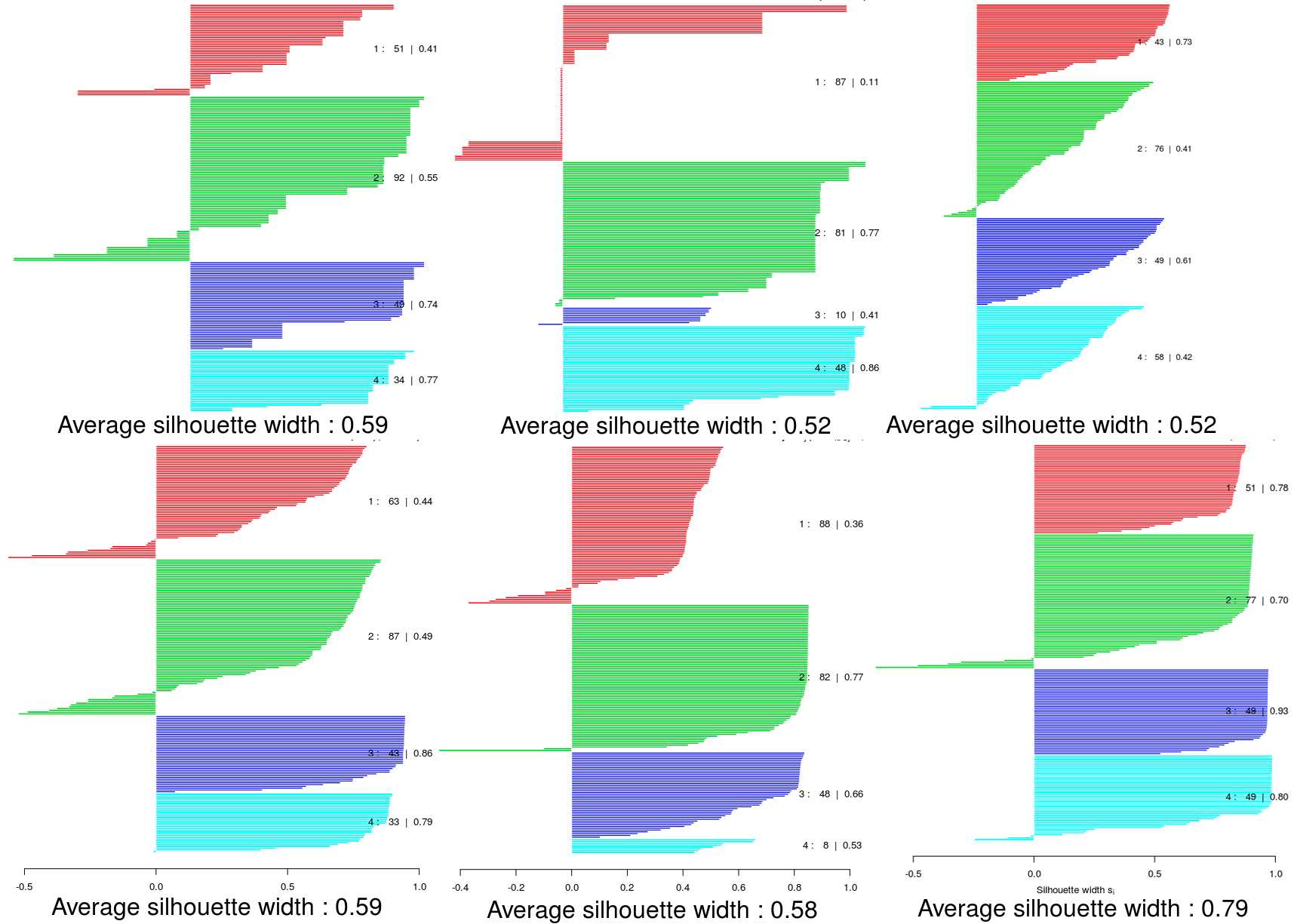}
\end{center}
%%\abovecaptionskip -3pt
\caption{\it Silhouette plot of K-means clustering, hierarchical clustering, SNF (top row) and one run of their consensus (bottom row).} 
\label{figure:fig3}
\end{figure}

For K-means clustering, hierarchical clustering, SNF, and their consensus, we evaluate their performance under two 
metrics---consensus matrix and ASW. The ensemble size in consensus are 10000; reducing to 500 incurs very little 
loss in accuracy (see simulations on the impact of ensemble size to results in the supporting material).
\\
\\
Figure~\ref{figure:fig2} shows the consensus matrix of consensus K-means clustering, hierarchical clustering, and SNF. 
Overall, all algorithms identify certain cluster structures in the data; all show a fairly clear block diagonal structure 
in their consensus matrix. Of the three consensus clustering algorithms, SNF does the best with a very clear block diagonal
structure, seconded by K-means clustering which identifies clusters agreeing well with those by SNF, and hierarchical 
clustering does poorly in identifying the clusters though being less noisy than K-means on off-diagonals. This is likely 
due to the flexibility of SNF in combining distances from different data and its use of spectral clustering which is effective 
in high dimensional data and in capturing various geometry in the data.
%%\\[-0.3in]
\begin{table}[htb]
\begin{center}
\begin{tabular}{c|c|c}
    \hline
~\textbf{Method}~                       & ~\textbf{Non-consensus}~                        &~\textbf{Consensus}~   \\
\hline 
K-means clustering                     &0.59               	&$0.5927 \pm 0.0113$         \\
Hierarchical clustering                &0.52              	&$0.5827 \pm 0.0236$          \\
SNF                                            &0.52                	&$0.7942 \pm 0.0093$          \\
\hline
\end{tabular}
\end{center}
\caption{\it{Average silhouette width for different clustering algorithms and their consensus. 
}} \label{table:avgASW}
\end{table}
%%\\
%%\\
\\
\\
Table~\ref{table:avgASW} shows the average silhouette width for different clustering algorithms and their consensus.
For consensus clustering, the results are averaged over 100 runs. 
Figure~\ref{figure:fig3} is the silhouette plot of the three non-consensus clustering algorithms and a typical run of their 
consensus. While consensus does not improve K-means clustering, it greatly improves hierarchical clustering and SNF
(ASW's improve so significantly that no statistical testing is necessary). Among non-consensus algorithms, 
K-means achieves the highest ASW, the ASW by hierarchical clustering and by SNF 
are similar but for different reasons (hierarchical clustering does poorly in identifying the clusters, while SNF and K-means 
clustering have a similar shape in the silhouette plot). For consensus algorithms, SNF outperforms K-means and hierarchical clustering by a large margin; this is consistent with
results shown in the consensus matrix. 
\\
\\
We also carry out additional experiments
on three more datasets, including the LUAD (Lung Adenocarcinoma) data, the UCEC (Uterine Corpus Endometrial Carcinoma) data,
and the LUSC (Lung Squamous Cell Carcinoma) data. Results of a similar pattern are observed. Additionally, we evaluate the effect 
of ensemble size to consensus clustering. We have tried ensemble size of 500, 5000, and 50000, and we find that this leads to very small
difference in consensus clustering results. Details can be seen at a supporting web (\url{https://sites.google.com/site/fsdm2018cwy2686/home}).

\section{Conclusions}
\label{section:conclusion}
We explored a data-driven approach in identifying cancer subtypes. In particular, we assess the suitability
of various clustering algorithms on cancer-omics data, including K-means clustering, hierarchical clustering, SNF and their consensus. 
Despite being vague on the number of clusters and the clustering assignment for a few observations, all non-consensus clustering
algorithms reveal useful information about cancer subtypes. Consensus clustering is very reliable in determining the number 
of clusters, and often outperforms non-consensus ones. Consensus of SNF achieves the best results in our study, with significant 
statistical confidence in terms of ASW and consensus matrix. We expect that more data, or more information
on the molecular mechanism that underlies the difference among subgroups, would further improve clustering and the associated 
statistical confidence.
\\
\\
In future work, we will explore two directions: to identify the differentially expressed genes, miRNA, and pathways, and to find the differences of clinical behaviors such as survival, drug and treatment sensitivity, and recurrence, among subgroups. We expect that this would help explain the difference in clinical behaviors. Such an understanding would benefit early cancer diagnosis and lead to more efficient treatments.

\bibliographystyle{plain}
%%\bibliography{myBib}

\begin{thebibliography}{10}

\bibitem{Bertino2002}
J.~Bertino.
\newblock {\em Encyclopedia of Cancer}.
\newblock Academic Press, New York, 2002.

\bibitem{who}
{World Health Organization}.
\newblock \url{http://www.who.int/news-room/fact-sheets/detail/cancer}.

\bibitem{nciCancer}
{US National Cancer Institute}.
\newblock
  \url{https://www.cancer.gov/about-cancer/understanding/what-is-cancer}.

\bibitem{DagogoShaw2018}
I.~Dagogo-Jack and A.~T. Shaw.
\newblock Tumour heterogeneity and resistance to cancer therapies.
\newblock {\em Nature Reviews Clinical Oncology}, 15:81--94, 2018.

\bibitem{RosenthalMcGranahan2017}
R.~Rosenthal, N.~McGranahan, J.~Herrero, and C.~Swanton.
\newblock Deciphering genetic intratumor heterogeneity and its impact on cancer
  evolution.
\newblock {\em Annual Review of Cancer Biology}, 1(1):223--240, 2017.


\bibitem{BijlsmaSadanandam2017}
M.~F. Bijlsma, A.~Sadanandam, P.~Tan, and L.~Vermeulen.
\newblock Molecular subtypes in cancers of the gastrointestinal tract.
\newblock {\em Nature Reviews Gastroenterology and Hepatology}, 14:333--342,
  2017.

\bibitem{KourouExarchos2015}
K.~Kouroua, T.~Exarchos, K.~Exarchos, M.~Karamouzis, and D.~Fotiadis.
\newblock Machine learning applications in cancer prognosis and prediction.
\newblock {\em Journal of Comp. and Structural Biotechnology}, 13:8--17, 2015.

\bibitem{Lyons2017ImmuneCP}
Y.~A. Lyons, S.~Wu, W.~W. Overwijk, K.~A. Baggerly, and A.~Sood.
\newblock Immune cell profiling in cancer: molecular approaches to
  cell-specific identification.
\newblock {\em npj Precision Oncology}, 1:1--8, 2017.


\bibitem{CamastraDTS2015}
F.~Camastra, M.~{Di Taranto}, and A.~Staiano.
\newblock Statistical and computational methods for genetic diseases: An
  overview.
\newblock {\em Computational and Mathematical Methods in Medicine}, 2015, 2015.

\bibitem{CGARN2014}
The Cancer Genome Atlas~Research Network.
\newblock Comprehensive molecular characterization of gastric adenocarcinoma.
\newblock {\em Nature}, 513:202--209, 2014.

\bibitem{TCGA2015}
K.~Tomczak, P.~Czerwinska, and M.~Wiznerowicz.
\newblock {The Cancer Genome Atlas (TCGA): an immeasurable source of
  knowledge}.
\newblock {\em Contemporary Oncology}, 19:A68--A77, 2015.

\bibitem{ZhangLiu2013}
W.~Zhang, Y.~Liu, N.~Sun, D.~Wang, X.~Dou J.~Boyd-Kirkup, and J.~Han.
\newblock Integrating genomic, epigenomic, and transcriptomic features reveals
  modular signatures underlying poor prognosis in ovarian cancer.
\newblock {\em Cell Reports}, 4:542--553, 2013.

\bibitem{MaZhang2017}
T.~Ma and A.~Zhang.
\newblock Integrate multi-omic data using affinity network fusion for cancer
  patient clustering.
\newblock In {\em IEEE International Conference on Bioinformatics and
  Biomedicine}, pages 398--403, 2017.

\bibitem{YangMichailidis2016}
Z.~Yang and G.~Michailidis.
\newblock A non-negative matrix factorization method for detecting modules in
  heterogeneous omics multi-modal data.
\newblock {\em Bioinformatics}, 32:1--8, 2016.

\bibitem{SNF2014}
B.~Wang, A.~Mezlini, F.~Demir, M.~Fiume, Z.~Tu, M.~Brudno, B.~Haibe-Kains, and
  A.~Goldenberg.
\newblock {Similarity network fusion for aggregating data types on a genomic
  scale}.
\newblock {\em Nature Methods}, 11:333--337, 2014.

\bibitem{MoWangSeshan2013}
Q.~Mo, S.~Wang, and V.~Seshan.
\newblock Pattern discovery and cancer gene identification in integrated cancer
  genomic data.
\newblock {\em Proceedings of the National Academy of Sciences, USA.},
  110(11):4245--4250, 2013.

\bibitem{lloyd1982}
S.~P. Lloyd.
\newblock Least squares quantization in {PCM}.
\newblock {\em IEEE Transactions on Information Theory}, 28(1):128--137, 1982.

\bibitem{KaufmanRousseeuw1990}
L.~Kaufman and P.~J. Rousseeuw.
\newblock {\em Finding Groups in Data: An Introduction to Cluster Analysis}.
\newblock Wiley, New York, 1990.

\bibitem{Ward1963}
J.~H. Ward.
\newblock Hierarchical grouping to optimize an objective function.
\newblock {\em Journal of the American Statistical Association},
  58(301):236--244, 1963.

\bibitem{ShiMalik2000}
J.~Shi and J.~Malik.
\newblock Normalized cuts and image segmentation.
\newblock {\em IEEE Transactions on Pattern Analysis and Machine Intelligence},
  22(8):888--905, 2000.


\bibitem{NgJordan2002}
A.~Y. Ng, M.~I. Jordan, and Y.~Weiss.
\newblock On spectral clustering: analysis and an algorithm.
\newblock In {\em Neural Information Processing Systems (NIPS)}, volume~14,
  2002.
  

\bibitem{StrehlGhosh2002}
A.~Strehl and J.~Ghosh.
\newblock Cluster ensembles -- a knowlwdge reuse framework for combining
  multiple partitions.
\newblock {\em Journal of Machine Learning Research}, 3:582--617, 2002.

\bibitem{MontiTamayo2003}
S.~Monti, P.~Tamayo, J.~Mesirov, and T.~Golub.
\newblock Consensus clustering: A resampling-based method for class discovery
  and visualization of gene expression microarray data.
\newblock {\em Machine Learning}, 52:91--118, 2003.

\bibitem{CF}
D.~Yan, A.~Chen, and M.~I. Jordan.
\newblock Cluster {F}orests.
\newblock {\em Computational Statistics and Data Analysis}, 66:178--192, 2013.





\bibitem{LiuShaoFu2016}
H.~Liu, M.~Shao, and Y.~Fu.
\newblock Consensus guided unsupervised feature selection.
\newblock In {\em Proceedings of the Thirtieth AAAI Conference on Artificial
  Intelligence}, 2016.

\bibitem{LiuShaoLiFu2017}
H.~Liu, M.~Shao, S.~Li, and Y.~Fu.
\newblock Infinite ensemble clustering.
\newblock {\em Data Mining and Knowledge Discovery}, 32(2):385--416, 2017.

\bibitem{WanAllenLiu2016}
Y.~W. Wan, G.~I. Allen, and Z.~Liu.
\newblock {TCGA2STAT: simple TCGA data access for integrated statistical
  analysis in R}.
\newblock {\em Bioinformatics}, 32(6):952--954, 2016.




\bibitem{Hotelling1933}
H.~Hotelling.
\newblock Analysis of a complex of statistical variables into principal
  components.
\newblock {\em Journal of Educational Psychology}, 24:417--441, 1933.




\bibitem{Rousseeuw1987}
P.~J. Rousseeuw.
\newblock Silhouettes: a graphical aid to the interpretation and validation of
  cluster analysis.
\newblock {\em Computational and Applied Mathematics}, 20:53--65, 1987.







\bibitem{YanHuangJordan2009}
D.~Yan, L.~Huang, and M.~I. Jordan.
\newblock Fast approximate spectral clustering.
\newblock In {\em Proceedings of the 15th ACM SIGKDD}, pages 907--916, 2009.

\bibitem{XingNgJordanRussell2002}
E.~P. Xing, A.~Y. Ng, M.~I. Jordan, and S.~Russell.
\newblock Distance metric learning, with application to clustering with
  side-information.
\newblock In {\em Proceedings of Neural Information Processing Systems (NIPS)},
  pages 521--528, 2002.


\end{thebibliography}

\end{document}